\documentclass[10pt, conference, compsocconf]{IEEEtran}
\usepackage{amsthm}
\usepackage{amssymb}
\usepackage{listings}

\newtheorem{definition}{Definition}
\usepackage{pgfplotstable}
\usepackage{pgfplots}
\usepackage{tabularx}
\usepackage{multirow}
\usepackage{graphicx}
\newcolumntype{R}{>{\raggedleft\arraybackslash}X}
\newcolumntype{L}{>{\raggedright\arraybackslash}X}
\newcolumntype{C}{>{\centering\arraybackslash}X}

\newcommand\Tstrut{\rule{0pt}{2.6ex}}

\usepackage{cite}

\ifCLASSINFOpdf
\else
\fi
\usepackage[cmex10]{amsmath}
\usepackage{algorithmic}
\begin{document}
%
\title{Heuristic algorithms for the Longest Filled Common Subsequence Problem}


\author{\IEEEauthorblockN{Radu Stefan Mincu}
\IEEEauthorblockA{Department of Computer Science\\
University of Bucharest\\
Bucharest, Romania\\
mincu.radu@fmi.unibuc.ro}
\and
\IEEEauthorblockN{Alexandru Popa}
\IEEEauthorblockA{Department of Computer Science\\
University of Bucharest\\
Bucharest, Romania\\
alexandru.popa@fmi.unibuc.ro}
\IEEEauthorblockA{
National Institute for Research \\ and Development in Informatics\\
Bucharest, Romania}

}


%


\maketitle

\begin{abstract}
At CPM 2017, Castelli et al. define and study a new variant of the Longest Common Subsequence Problem, termed the Longest Filled Common Subsequence Problem (LFCS). For the LFCS problem, the input consists of two strings $A$ and $B$ and a multiset of characters $\mathcal{M}$. The goal is to insert the characters from $\mathcal{M}$ into the string $B$, thus obtaining  a new string $B^*$, such that the Longest Common Subsequence (LCS) between $A$ and  $B^*$ is maximized. Casteli et al. show that the problem is NP-hard and provide a 3/5-approximation algorithm for the problem. 

In this paper we study the problem from the experimental point of view. We introduce, implement and test new heuristic algorithms and compare them with the approximation algorithm of Casteli et al. Moreover, we introduce an Integer Linear Program (ILP) model for the problem and we use the state of the art ILP solver, Gurobi, to obtain exact solution for moderate sized instances.


\end{abstract}


%
\IEEEpeerreviewmaketitle

\section{Introduction}

\paragraph*{\sc Motivation and previous work}

The Longest Common Subsequence problem (LCS) is long studied~\cite{Apostolico1997,Bergroth2000,Hirschberg1977AlgorithmsFT} and its importance is further emphasized by its a wide array of applications in diverse fields such as data compression, computational biology, text editing and comparison (notable examples include the Unix \lstinline|diff| utility and plagiarism detection systems), pattern recognition.

In the field of computational biology, advances in genome sequencing techniques are bringing about the need for algorithms to be used in the analysis and reconstruction of genomic data. As a consequence, many LCS variants have been developed to interpret the DNA sequence fragments resulting from various DNA sequencing procedures for the purpose of reconstructing a genome. Among these studied problems is the \emph{Constrained LCS problem (CLCS)} first presented by Tsai \cite{TSAI2003173} which is proven to be equivalent to a particular case of \emph{Constrained Sequence Alignment (CSA)} \cite{CHIN2004175}. 
There are applications for the LCS problem in comparing genome data such as the exemplar model based LCS variants, namely the \emph{Exemplar LCS (ELCS)} variants~\cite{Bonizzoni2007} and the \emph{Repetition Free LCS (RFLCS)} problem~\cite{ADI20101315}. A generalization of the CLCS and RFLCS problems exists in the \emph{Doubly Constrained LCS problem (DCLCS)}~\cite{BONIZZONI2010877}. 

In CPM 2017, Castelli, Dondi, Mauro and Zoppis introduce a new LCS variant called \emph{Longest Filled Common Subsequence (LFCS)}~\cite{castelli2017} inspired from the particulars of the \emph{Scaffold Filling problem} in genome reconstruction~\cite{Munoz2010,BULTEAU201572}.  The LFCS problem is defined as follows:

\begin{definition}[Longest Filled Common Subsequence] Let there be two strings $A$ and $B$ over alphabet $\Sigma$ and a multiset $\mathcal{M}$ of symbols in $\Sigma$. A \emph{filling} $B^*$ of string $B$ is defined as a sequence obtained from $B$ by inserting a subset of the symbols from $\mathcal{M}$ into $B$. Find a filling $B^*$ that maximizes the length of the longest common subsequence of strings $A$ and $B^*$. 
\end{definition}

Castelli et al. prove that LFCS is $\mathcal{APX}$-hard even when $A$ contains at most two occurrences of each symbol in $\Sigma$~\cite{castelli2017}. They also show a 3/5-approximation algorithm and a fixed parameter algorithm where the parameter is the number of symbols from $\mathcal{M}$ inserted in the string $B$.

In this paper we study the problem from the experimental point of view. We introduce, implement and test new heuristic algorithms and compare them with the approximation algorithm of Casteli et al. Moreover, we give an ILP formulation for the problem and we use the state of the art ILP solver, Gurobi, to obtain exact solution for moderate sized instances.

The rest of the paper is organized as follows.

In Section \ref{sec:prel} we describe how to reduce the search space for LFCS and how to obtain bounds for the solution value. Section \ref{sec:ILP} contains an ILP model used to obtain optimum solutions. In practice, exact methods such as ILP solving might take a long time to complete, therefore we construct heuristic algorithms in Section \ref{sec:heuristics} and test them on procedurally generated data. The results and the experiments are described in Section \ref{sec:experiments}.




\section{Preliminaries}
\label{sec:prel}

\subsection{LFCS Solution Space}

We wish to reason about the solution space of the LFCS problem to offer some insights into its nature. To begin with, we may consider the extreme where $\mathcal{M} =\emptyset$. Then, the solution for LFCS is the same as the LCS value of inputs $A$ and $B$. Conversely, if $\mathcal{M}$ contains $A$ entirely, then we can just pick as a filling $B^*=AB$, and the solution for LFCS is of size $|A|$. Clearly, the difficult cases to solve are when $\mathcal{M}$ is in the middle-ground. If $\mathcal{M}$ is very small, then there are few insertions to consider. Moreover, if $\mathcal{M}$ is too extensive, then the same situation occurs. To better understand the solution space, let us consider an alternative formulation of LFCS:

\begin{definition}[Longest Filled Common Subsequence (alt.)] Let there be two strings $A$ and $B$ over alphabet $\Sigma$ and a multiset $\mathcal{M}$ of symbols in $\Sigma$. Let $A'$ be string $A$ from which we have removed a subset of symbols, no more than the amount contained in $\mathcal{M}$. Find $A'$ that maximizes the sum between $|A|-|A'|$ and the length of the longest common subsequence of strings $A'$ and $B$. 
\end{definition}

The above definition changes the focus from inserting elements from $\mathcal{M}$ into $B$ to deleting the elements matched with $\mathcal{M}$ from $A$. The reason why we prefer the latter formulation is because the search space is smaller and much easier to compute than the former, while the optimum value of the problem remains the same.

\begin{figure}[!h]
\begin{center}
\includegraphics[width=0.4\linewidth]{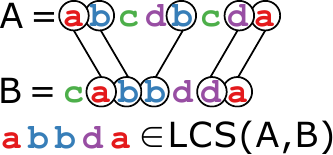}
\end{center}

\includegraphics[width=0.49\linewidth]{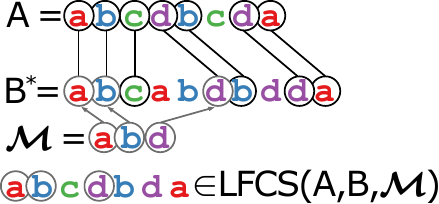}
\includegraphics[width=0.49\linewidth]{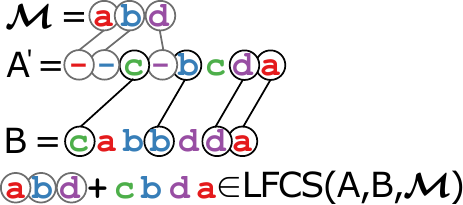}

\caption{Examples of LCS and LFCS. On the top side is one LCS of $A$, $B$. On the bottom side, \lstinline|abd| are matched with $\mathcal{M}$, and \lstinline|cbda| are LCS-matched. Standard LFCS is used on the left side, while the alternative LFCS formulation is illustrated on the right side.}
\label{fig:lcs}
\end{figure}

We illustrate this with an example in Figure \ref{fig:lcs}: \lstinline|abd| are matched with $\mathcal{M}$, while \lstinline|cbda| is the LCS of $A'=$\lstinline|cbcda| and $B=$\lstinline|cabbdda|. Alternatively, we may consider the LCS of $A$ with one of the optimum \lstinline|$B^*\ni\{$abcdabbdda$\text{, }$abcadbbdda$\text{, }$abcabdbdda$\}$|. Thus, it is equivalent to say that \lstinline|d| can be inserted into three places in $B$ or to say that \lstinline|d| from $A$ is matched with $\mathcal{M}$ for the purpose of solving the problem.

To obtain solutions for LFCS, we have seen that we can focus on deleting the symbols from $A$ that are matched with $\mathcal{M}$, effectively reducing the search space. But it is possible to do even better.

\paragraph{Observation 1} Notice that we may only match symbols of $A$ with as many symbols as there exist in $\mathcal{M}$.
\paragraph{Observation 2} Given an optimum solution for LFCS, it either has the maximum number of matched symbols with $\mathcal{M}$ or one may construct an optimum solution for LFCS that uses an extra symbol from $\mathcal{M}$ by picking a symbol present in both the LCS and in the unmatched subset of $\mathcal{M}$ and matching it with $\mathcal{M}$.

Using the above observations we may compute the search space for a given instance by counting all $A'$. Observations 1,2 help us to reduce the search space further by allowing us to disregard all $A'$ that do not have the maximum number of matchings with $\mathcal{M}$. As such, in order to describe the search space, one may enumerate all possible ways to match the maximum number of symbols in $\mathcal{M}$ with the symbols of $A$.

Going back to the example in Figure \ref{fig:lcs}, we wish to compute the size of the search space $S(A,\mathcal{M})$. We see that we can match \lstinline|d|, \lstinline|a| and \lstinline|b| each in two places in $A$ , yielding a total of 8 combinations. In the general case, we go though $\mathcal{M}$ and for each distinct symbol we calculate all the possible ways to match the occurrences in $A$ with exactly the amount available in $\mathcal{M}$. If we do this for each symbol, we get a product of combinations. More formally:

Let  $a(\sigma)=\lvert A \rvert_{\sigma}$, $m(\sigma)= \min(\lvert \mathcal{M} \rvert_{\sigma},\lvert A \rvert_{\sigma})$. Then:
\begin{equation}
S(A,\mathcal{M})=\prod_{\sigma \in \Sigma} { a(\sigma) \choose m(\sigma)}
\end{equation}

The ability to compute the size of the search space becomes important when we consider heuristic algorithms based on randomization. For example, if the search space is reasonably small, we may obtain an optimum solution by repeatedly evaluating random solutions.

\subsection{Bounds for the LFCS Solution}

As a lower bound for the LFCS problem is first given by the LCS of the $A$, $B$ inputs. For a tighter and quite practical lower bound we have used in our experiments the 3/5-approximation algorithm from \cite{castelli2017}.

When considering an upper bound for the LFCS solution, one quick method is to sum the value of LCS($A$,$B$) with \mbox{$\lvert A \cap \mathcal{M} \rvert$} (as a multiset intersection). In other words, the value of the solution cannot exceed the case where the symbols in $A$ are maximally matched with $\mathcal{M}$ and the sets of symbols matched with $\mathcal{M}$ and by LCS are disjoint.
\section{ILP model for the LFCS problem}
\label{sec:ILP}

In this section we show an ILP formulation for the LFCS problem. Let $n = |A|$ and $m = |B|$. We define a variable $x_{ij}$ for every $i \in \{1, \dots, n\}$ and $j \in \{1, \dots, m \}$. We have $x_{ij} = 1$ if $a_i = b_j$ and these two charcters belong to the LFCS solution. We also define a variable $y_i$ for every $i \in \{1, \dots, n\}$ which is equal to $1$ if $a_i$ is matched with a character from $\mathcal{M}$. The full ILP model follows:
\begin{subequations}
\begin{align}
\max  \sum_{i=1}^n y_i + \sum_{i=1}^n \sum_{j=1}^m x_{ij}  & \text{ subject to:}\\
x_{ij} + x_{kl} \le 1 & \text{, } \forall (i<k) \wedge (j>l) \label{eq:cross}\\
y_i + \sum_{j=1}^m x_{ij} \le 1 & \text{, }\forall i \in \{1,\dots,n\}\label{eq:one}\\
\sum_{i=1}^n x_{ij} \le 1 & \text{, }\forall j \in \{1,\dots,m\} \label{eq:two}\\
\sum_{i=1,a_i=\sigma}^n y_{i} \le \lvert \mathcal{M}\rvert_{\sigma} & \text{, }\forall \sigma \in \mathcal{M} \label{eq:count}\\
x_{ij}, y_i  & \in \{0,1\}
\end{align}
\end{subequations}

Explanation for the constraints in the ILP:
\begin{enumerate}
\item Constraint \ref{eq:cross} ensures that two pairs of characters that match are non-crossing.
\item Constraint \ref{eq:one} ensures that each character from $A$ may only be matched with a character from $\mathcal{M}$ or a single character from $B$.
\item Constraint \ref{eq:two} ensures that each character from $B$ may only be matched with a  single character from $A$.
\item Constraint \ref{eq:count} ensures that the characters from $A$ that are matched with $\mathcal{M}$ do not exceed the total amount available in $\mathcal{M}$ for that particular character.
\end{enumerate}

\section{Heuristic algorithms for the LFCS problem}
\label{sec:heuristics}

In this section we describe heuristics for obtaining solutions for LFCS in a practical setting. 

\subsection{Uniformly Sampled Solutions}

A straightforward method to approach the problem is to solve uniformly random solutions in the search space of a given instance. The solutions are determined by selecting a random combination of symbols from $A$ to match with $\mathcal{M}$, such that $A$ is maximally matched with $\mathcal{M}$. This can be achieved procedurally by uniform sampling of the index-set of each symbol in the alphabet. Subsequently, an LCS is performed on the remainder $A'$ and $B$ and the value of the LFCS solution will be LCS($A'$,$B$) + \mbox{$\lvert A \cap \mathcal{M} \rvert$}. For our experiments we have chosen to generate a constant 10000 random solutions. For more consistency, it may be advisable for the sample count to be a fraction of  the search space (as computed in Section \ref{sec:prel}).

\subsection{Local Search Algorithm}

We consider an algorithm inspired from local search techniques and greedy hillclimbing. We start with $A$ as an initial solution and wish to construct $A'$ by matching symbols of $A$ with $\mathcal{M}$. 

\begin{figure}
\includegraphics[width=\linewidth]{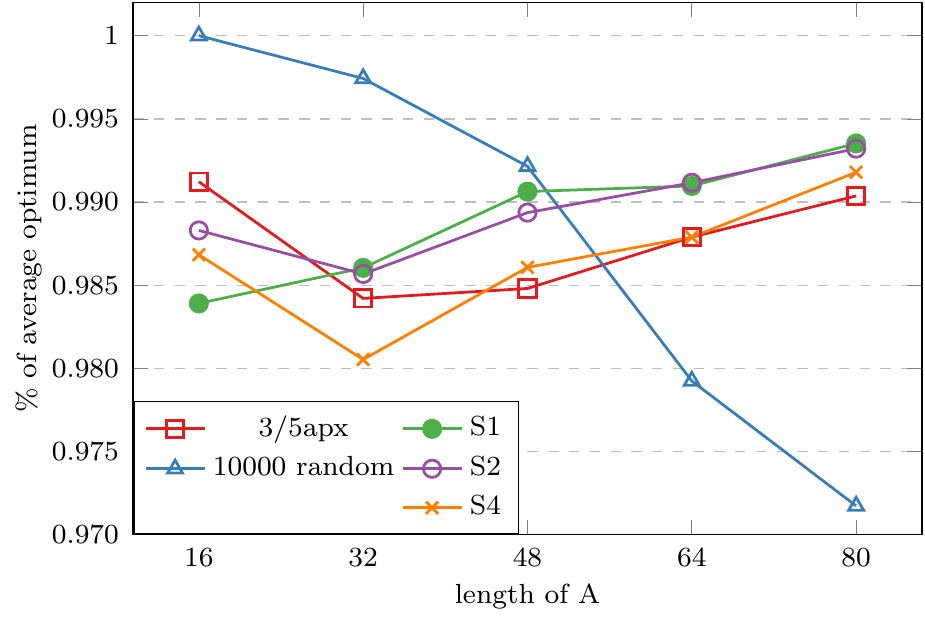}
\caption{Average of the solutions returned by the algorithms. All values are divided by the average optimum. The alphabet of the instances is of size 1/8 the length of A.}
\label{fig:mysuperultraniceplothaha1}
\end{figure}
\begin{figure}
\includegraphics[width=\linewidth]{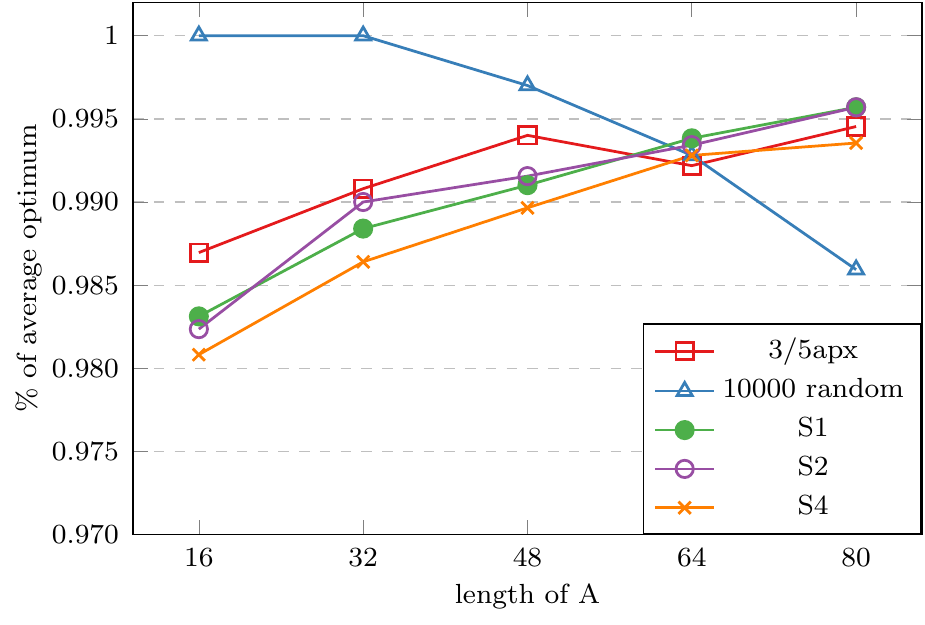}
\caption{Average of the solutions returned by the algorithms. All values are divided by the average optimum. The alphabet of the instances is of size 1/4 the length of A.}
\label{fig:mysuperultraniceplothaha2}
\end{figure}
\begin{figure}
\includegraphics[width=\linewidth]{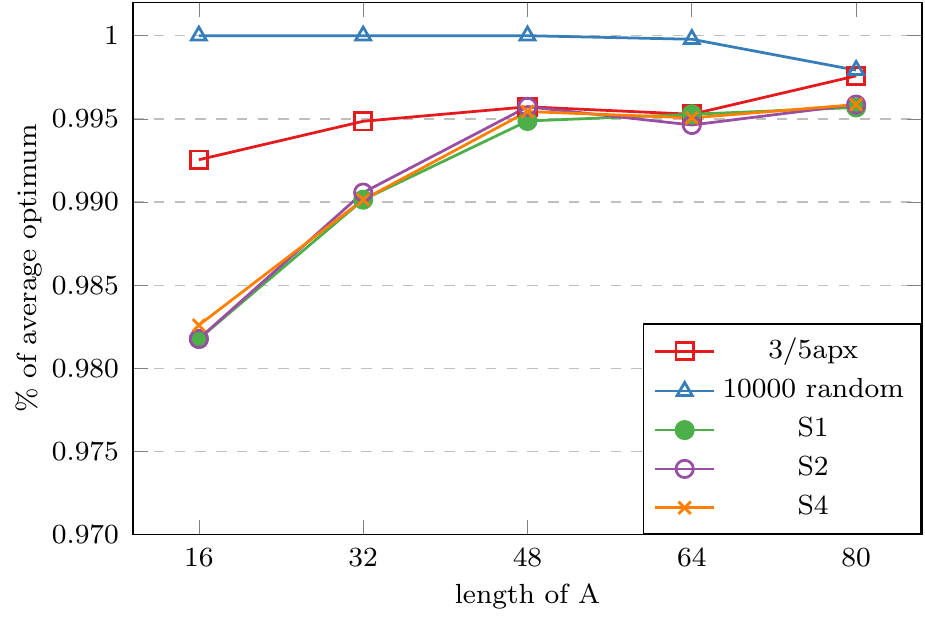}
\caption{Average of the solutions returned by the algorithms. All values are divided by the average optimum. The alphabet of the instances is of size 1/2 the length of A.}
\label{fig:mysuperultraniceplothaha3}
\end{figure}

At each iteration of the algorithm we wish to match with $\mathcal{M}$ (i.e. delete) as many as $k$ symbols, located on $k$ consecutive positions in $A$, such that the sum of the number of matched (i.e. deleted) symbols and the LCS of the unmatched remainder of $A$ and $B$ is the maximum. The best solution becomes the new incumbent solution. The algorithm terminates when we fail to match at least 1 new symbol.

For each incumbent solution $I$, the neighborhood to be explored is composed of all strings $I'$ such that $I'$ is obtained by deleting symbols within any window of length $k$ from $I$. Here, \textit{window} means exactly $k$ consecutive positions in $I$. In other words, we start with a left-to-right sliding window of length $k$ and attempt all $2^k$ possible ways to match the symbols, if the amount in $\mathcal{M}$ is sufficient. If $\lvert A \rvert =n$, we compute at most $2^k (n-k)$ LCS-s at each iteration.

We dub our algorithm S$k$, where the parameter $k$ is the length of the window.

\section{Experimental results} 
\label{sec:experiments}

We run our algorithms and the 3/5-approximation from \cite{castelli2017} on a selection of instances
that are randomly and procedurally generated. We find the optimum solution for each instance using an implementation of our ILP model in Gurobi/Python.

To obtain each $(A,B,\mathcal{M})$ instance we first generate a uniform string for $A$. To make $B$, we first copy $A$ and iterate through it, changing symbols with a probability of 50\%. The change applied is uniformly chosen among duplication, deletion and substitution with a different symbol. We then randomly split $B$ into segments no longer than $|A|/8$ and discard $>$30\% of them. The remainder is the final $B$ string, while the discarded symbols are put into $\mathcal{M}$.

For our tests we select $n= \lvert A \rvert$ to be 16, 32, 48, 64, 80, while the alphabet size is $\frac{n}{8}$, $\frac{n}{4}$, $\frac{n}{2}$. For each combination of $n$ and $\lvert \Sigma \rvert$, we generate 100 instances.

The results are showcased in Table \ref{table:mysuperultranicetablehaha} and Figures \ref{fig:mysuperultraniceplothaha1}, \ref{fig:mysuperultraniceplothaha2}, \ref{fig:mysuperultraniceplothaha3}. All the algorithms arrive within 97\% of the optimum. We can observe that the 3/5-approximation algorithm is quite good in the general case, but there are instances where our local search algorithm S$k$ outperforms the 3/5-approximation algorithm. This is noticeable for small alphabets and in Figure \ref{fig:mysuperultraniceplothaha1}, where all S$k$ perform better for lengths 48, 64, 80. As expected, random sampling for solutions shines when the solution space is smaller, which is the case for Figure \ref{fig:mysuperultraniceplothaha3}.

Because the choices of algorithm S$k$ are not reversible, the algorithm is prone to get stuck in local optima as $k$ increases. To improve the clarity of the figures, we limited our selection to S1, S2, S4, which are among the best tested.

\begin{table}
\caption{Table illustrating how many times each algorithm reaches an optimum solution in our experiments on 100 instances (each case). The first column describes the alphabet size, then follows the length of the A input strings and then are how many optima each algorithm finds.}
\label{table:mysuperultranicetablehaha}
\begin{center}
\begin{tabular}{  | c| c| c| c| c| c| c|}
\hline
\Tstrut
$\Sigma$ & n & 3/5apx & rand & S1 & S2 & S4\\
\hline
\Tstrut
\multirow{5}{*}{$\dfrac{n}{8}$} & 16 & 88 & 100 & 78 & 84 & 82\\
 \cline{2-7}
 \Tstrut
 & 32 & 66 & 93 & 68 & 65 & 56\\
 \cline{2-7}
 \Tstrut
 & 48 & 53 & 70 & 68 & 63 & 54\\
 \cline{2-7}
 \Tstrut
 & 64 & 50 & 22 & 57 & 60 & 51\\
 \cline{2-7}
 \Tstrut
 & 80 & 53 & 9 & 65 & 62 & 57\\
 \hline
 \Tstrut
\multirow{5}{*}{$\dfrac{n}{4}$} & 16 & 83 & 100 & 78 & 77 & 77\\
 \cline{2-7}
 \Tstrut
 & 32 & 77 & 100 & 72 & 75 & 68\\
 \cline{2-7}
 \Tstrut
 & 48 & 80 & 89 & 71 & 71 & 70\\
 \cline{2-7}
 \Tstrut
 & 64 & 70 & 73 & 73 & 72 & 69\\
 \cline{2-7}
 \Tstrut
 & 80 & 71 & 36 & 78 & 79 & 67\\
 \hline
 \Tstrut
\multirow{5}{*}{$\dfrac{n}{2}$} & 16 & 91 & 100 & 79 & 80 & 80\\
 \cline{2-7}
 \Tstrut
 & 32 & 89 & 100 & 77 & 78 & 77\\
 \cline{2-7}
 \Tstrut
 & 48 & 85 & 100 & 82 & 86 & 85\\
 \cline{2-7}
 \Tstrut
 & 64 & 79 & 99 & 79 & 76 & 77\\
 \cline{2-7}
 \Tstrut
 & 80 & 87 & 90 & 76 & 77 & 76\\
 \hline
\end{tabular}
\end{center}
\end{table}

\section{Conclusions and future work}
In this paper study the LFCS problem introduced by Casteli et al. from an experimental point of view. We give  ILP formulations for the problem and we use the state of the art ILP solver, Gurobi, to obtain exact solution for moderate sized instances.  We introduce, implement and test new heuristic algorithms and compare them with the approximation algorithm of Casteli et al..

A natural open question is to find a better than 3/5  approximation algorithm for the LFCS problem.

\bibliographystyle{IEEEtran}
\bibliography{bibliography}
%



\end{document}